\documentclass[a4paper]{article}

\usepackage[pdftex]{graphicx}
\DeclareGraphicsExtensions{.pdf}

\usepackage{a4wide, times}

\usepackage[caption=false,font=footnotesize]{subfig}
\usepackage{amsmath}

\usepackage[latin1]{inputenc}
\usepackage[ngerman, english]{babel}

\usepackage{authblk}

\usepackage{color}

\usepackage{fancyhdr}

\newcommand{\yP}{\dot{y}}
\newcommand{\yPP}{\ddot{y}}

\newcommand{\omP}{\dot{\omega}}

\newcommand{\xP}{\dot{x}}

\newcommand{\xhat}{\widehat{x}}
\newcommand{\yhat}{\widehat{y}}
\newcommand{\zhat}{\widehat{z}}

\newcommand{\zhatB}{\widehat{\B{z}}}

\newcommand{\xhatP}{\dot{\widehat{x}}}

\newcommand{\x}{\times}

\newcommand{\DS}{\displaystyle}

\newcommand{\SStyle}{\scriptstyle}

\newcommand{\Durch}[2]{\frac{\DS #1}{\DS #2}}

\newcommand{\B}[1]{\mathbf{#1}}

\newcommand{\fmax}{\overline{f}}
\newcommand{\fmin}{\underline{f}}

\newcommand{\rads}{\frac{\text{rad}}{\text{s}}}

\newcommand{\regsym}{\textsuperscript{\textregistered}}

\definecolor{DarkGreen}{rgb}{0, 0.7, 0.2}

\allowdisplaybreaks[1]

\title{Wind Turbine Model and Observer in Takagi-Sugeno Model Structure}

\date{}

\author[1]{S\"oren Georg\thanks{soeren.georg@htw-berlin.de}}
\author[1,2]{Matthias M\"uller}
\author[1]{Horst Schulte\thanks{horst.schulte@htw-berlin.de}}

\affil[1]{HTW Berlin, Department of Engineering I, Control Engineering, Berlin, Germany}
\affil[2]{Key Wind Energy GmbH, Berlin, Germany}

\begin{document}

\maketitle

\thispagestyle{fancy}
\lhead{This is a preprint version of a paper that is to be included in the proceedings of the EAWE conference \\ "The Science of Making Torque from Wind", Oldenburg, Germany, October 2012, due to be published in the IOP Journal of Physics: Conference Series (JPCS) - in press.}

\begin{abstract}
Based on a reduced-order, dynamic nonlinear wind turbine model in Takagi-Sugeno (TS) model structure, a TS state observer is designed as a disturbance observer to estimate the unknown effective wind speed. The TS observer model is an exact representation of the underlying nonlinear model, obtained by means of the sector-nonlinearity approach. The observer gain matrices are obtained by means of a linear matrix inequality (LMI) design approach for optimal fuzzy control, where weighting matrices for the individual system states and outputs are included.
The observer is tested in simulations with the aero-elastic code FAST for the NREL 5 MW reference turbine, where it shows a stable behaviour both for IEC wind gusts and turbulent wind input.
\end{abstract}

\section{Introduction}

Takagi-Sugeno (TS) models provide a useful and uniform framework for nonlinear controller and observer design for dynamic systems. Originally introduced in the context of fuzzy systems \cite{TakagiSugeno}, TS models are weighted combinations of linear submodels and can either be derived from input-output data via system identification \cite{TakagiSugeno, SugenoKang:1988} or from mathematical models of nonlinear systems. Methods based on solving linear matrix inequalities (LMIs) allow for implicit stable controller and observer design for TS models \cite{WangTanakaGriffin:1996, Tanaka:2001, Lendek:2010}.

In this paper, a TS observer is designed as a disturbance observer to estimate the unknown effective wind speed from the available measurable system outputs. This observer is intended as a module for a fault-tolerant control scheme for wind turbines, where a reliable wind speed estimate is important, both as an input signal for fault-detection and isolation units (FDI) and as a redundant wind speed signal for the supervisory wind turbine control system.

Other methods have been applied to wind speed estimation in the literature. See for example \cite{MaPoulsen_EstWind:1995}, where Kalman filtering, extended Kalman filtering and the Newton-Raphson method are used and compared. Other dedicated algorithms have been applied, too. In \cite{Ostergaard_v_est:2007}, a state-observer for the rotor speed is combined with a PI controller to estimate the aerodynamic rotor torque. The effective wind speed is then reconstructed from the estimated torque signal via inversion of the aerodynamic model.
While being able to yield good wind speed estimates, these methods also have certain detriments.
The Kalman filter is only applicable to linear state-space models. Thus, estimating the wind speed for a wind turbine using a Kalman filter works only in the region of one operating point of a linearised wind turbine model. A possible remedy is provided by the extended Kalman filter, however, it is not possible to verify formal stability for the error dynamics, since the extended Kalman filter is an adaptive method.
\newline For observers in TS structure, however, the formal stability of the error dynamics can, at least in principle, be shown using linear matrix inequalities (LMI). The TS observer structure can also be extended to a TS sliding mode observer used for fault estimation \cite{GerlandSchulte:2010}. This has been applied in \cite{SchulteZajac:2012} for unknown load estimation and sensor fault reconstruction in pitch systems of wind turbines. For these reasons, and to achieve a certain level of uniformity within the design methods for different modules of a fault-tolerant control scheme, an observer in TS structure is used here for the estimation of the effective wind speed.

This paper is organised as follows. In section \ref{Sec_WTModel}, the reduced-order wind turbine model that serves as a basis for the observer is introduced.
In section \ref{Sec_TSObs}, the TS model structure is introduced along with an illustrating example. The observer is derived in TS structure and the method to obtain the observer gain matrices is discussed. Simulation results are presented in section \ref{Sec_Simulation}.

\section{\label{Sec_WTModel}Wind Turbine Model}

For the purpose of model-based control design, reduced-order models like those in \cite{Bindner:1999, Bianchi:2007} are appropriate, since they capture only the dominant system dynamics that are directly influenced by the control action \cite{Bianchi:2007}. A reduced-order model inspired from \cite{Bianchi:2007}, which was derived in TS structure in \cite{Georg:Fuzz2012}, is briefly introduced in this section and serves as a basis for the observer design in section \ref{Sec_TSObs}. In order to test the observer with a more realistic wind turbine model, the aero-elastic code FAST by NREL \cite{FASTUserGuide} is used for the simulation studies (see section \ref{Sec_Simulation}).
\newline Four degrees of freedom are considered for the reduced-order model: rotor and generator rotation angles ($\theta_r$, $\theta_g$), fore-aft tower top deflection $y_T$ and flapwise blade tip deflection $y_B$. The equations of motion, which describe the dynamics of the mechanical model, are obtained as

\begin{eqnarray}
	\left(m_T \, +\, N m_B\right) \, \yPP_T \, + \, N m_B \, \yPP_B \, + \, d_T \, \yP_T \, + \, k_T \, y_T  	& = & F_T	\label{Eq_EOM_Mech1} \\
	N m_B \, \yPP_T \, + \, N m_B \, \yPP_B \, + \, N d_B \, \yP_B \, + \, N k_B \, y_B 						& = & F_T	\label{Eq_EOM_Mech2} \\
	J_r \,\omP_r \, + \, d_S \, \left(\omega_r \, -\, \omega_g\right) \, + \, k_S \, \theta_s 					& = & T_a	\label{Eq_EOM_Mech3} \\
	J_g \,\omP_g \, - \, d_S \, \left(\omega_r \, -\, \omega_g\right) \, - \, k_S \, \theta_s 					& = & -T_g	\label{Eq_EOM_Mech4} \, ,
\end{eqnarray}
\newline where $N$ denotes the number of rotor blades, $R$ the rotor radius, $m_T$ and $m_B$ the effective tower and blade masses, $k_T$ and $k_B$ the effective stiffness coefficients for the tower top and blade tip deflection, $d_T$ and $d_B$ the damping coefficients for the respective tower and blade dynamics. $\theta_s = \theta_r - \theta_g$ denotes the shaft torsion angle, $T_a$ the aerodynamic rotor torque and $T_g$ the applied generator torque. An ideal gearbox is assumed, where the gearbox ratio is set to 1 for reasons of simplicity.
\newline Due to centrifugal forces acting on the rotor blades, the structural blade stiffness parameter can be modified by a term dependent on rotor angular velocity:

\begin{equation}
k_{B,\text{eff}} \, = \, k_B \, + \, k_B^\text{centr}\left(\omega_r\right) \, = \, k_B\, + \, \alpha \, m_B \, r_B \, \omega_r^2 \, ,
\label{Eq_kBEff}
\end{equation}
\newline where $r_B$ denotes the distance from the blade root to the blade centre of mass and $\alpha$ is a constant that needs to be adjusted to the simulated turbine. The inclusion of a centrifugal term is inspired from the FAST simulation software, where the correction is done for every blade section.

The pitch dynamics can be incorporated into the wind turbine model as a first-order delay model, $\, \tau_\beta\,\dot{\beta} + \beta = \beta_d\,$, where $\beta_d$ denotes the demanded pitch angle and $\tau_\beta$ the delay time constant.
\newline Introducing the state vector
$\B{x} \, = \, \left(y_T \quad y_B \quad \theta_s \quad \yP_T \quad \yP_B \quad \omega_r \quad \omega_g \quad \beta \right)^T$ and the input vector
$\B{u} \, = \, \left(\beta_d \quad T_g\right)^T$ , the system of dynamic
equations \eqref{Eq_EOM_Mech1} to \eqref{Eq_EOM_Mech4} including the centrifugal term \eqref{Eq_kBEff}
and the pitch dynamics can be transformed to state-space form:

\begin{align}
{\SStyle \xP_1} & = \, {\SStyle x_4} & &																										\label{Eqn_SSMdl_1} \\
{\SStyle \xP_2} & = \, {\SStyle x_5} & &																										\label{Eqn_SSMdl_2}	\\
{\SStyle \xP_3} & = \, {\SStyle x_6 - x_7} & &																									\label{Eqn_SSMdl_3}	\\
{\SStyle \xP_4} & = \, {\SStyle \frac{1}{m_T} \left(- \, k_T \, x_1 \, + \, N \, k_B \,x_2 	\, - \, d_T \, x_4 \, + \, N \, d_B \, x_5\right)}
									& & \, {\SStyle + \, \frac{N}{m_T} \, k_B^\text{centr}\left(\omega_r\right) \, x_2} 						\label{Eqn_SSMdl_4}	\\
{\SStyle \xP_5} & = \, {\SStyle \frac{k_T}{m_T} \, x_1 \, - \, \frac{m_T + N m_B}{m_B \,m_T} \, k_B \,x_2 \,
						+ \, \frac{d_T}{m_T}\,x_4 \,  -  \, \left(\frac{1}{m_B} + \frac{N}{m_T}\right)d_B\,x_5 \,}
						& & \, {\SStyle + \, \frac{m_T + N m_B}{m_B \,m_T} \, k_B^\text{centr}\left(\omega_r\right) \,x_2 \,
							+ \, \frac{1}{N m_B}\, F_T	}																						\label{Eqn_SSMdl_5} \\
{\SStyle \xP_6} & = \, {\SStyle -\frac{1}{J_r} \, \left( d_S \, \left(x_6 \, - x_7 \right) \, + \, k_S \, x_3 \right) \,}
						& & \, {\SStyle + \, \frac{1}{J_r} \,T_a}																				\label{Eqn_SSMdl_6} \\
{\SStyle \xP_7} & = \, {\SStyle \frac{1}{J_g} \, \left( d_S \, \left(x_6 - x_7\right) \,
						+ \, k_S \, x_3 \right) \,} & \, {\SStyle - \, \frac{1}{J_g}\,u_2} &													\label{Eqn_SSMdl_7} \\
{\SStyle \xP_8} & = \, {\SStyle -\frac{1}{\tau}\,x_8 \,} & {\SStyle + \, \frac{1}{\tau}\,u_1} &  \, ,											\label{Eqn_SSMdl_8}
\intertext{which can also be written in matrix form as}
 \B{\xP} 		& = \, \B{A \, x} \, 	& \, + \, \B{B \, u} \, & + \, \B{g}(\B{x},v) \, ,														\label{Eqn_StateSpaceMdl}
\end{align}
\newline with system matrix $\B{A}$, input matrix $\B{B}$ and a nonlinear state vector $\B{g}(\B{x},v)$.
\newline The aerodynamic rotor thrust and torque are given by $F_T = \frac{\rho \pi R^2}{2} \, C_T\left(\lambda,\beta\right) \, v^2$ and
\newline $T_a	= \frac{\rho \pi R^3}{2} \, C_Q\left(\lambda,\beta\right) \, v^2$, where $R$ denotes the rotor radius, $\rho$ the air density,
$v$ the wind speed and $\lambda = R \,\frac{\omega_r}{v}$ the tip speed ratio. $C_Q$ and $C_T$ are the aero maps for the rotor thrust and torque coefficients.
Due to the expressions for $F_T$ and $T_a$, the state-space model \eqref{Eqn_StateSpaceMdl} is nonlinear.

\subsection{\label{Sec_ModelParameters}Model Parameters}

The model parameters for the turbine model \eqref{Eqn_StateSpaceMdl} are based on the NREL 5 MW reference turbine \cite{Jonkman:2009}. The parameters are listed in \ref{Sec_AppModelParameters}, some of which can be directly taken from \cite{Jonkman:2009} or example input and log files of FAST simulation runs of the 5 MW reference turbine.

\subsubsection{\label{Sec_StructParameters}Structural Parameters}

\begin{figure}
\centering
\includegraphics[scale=0.42]{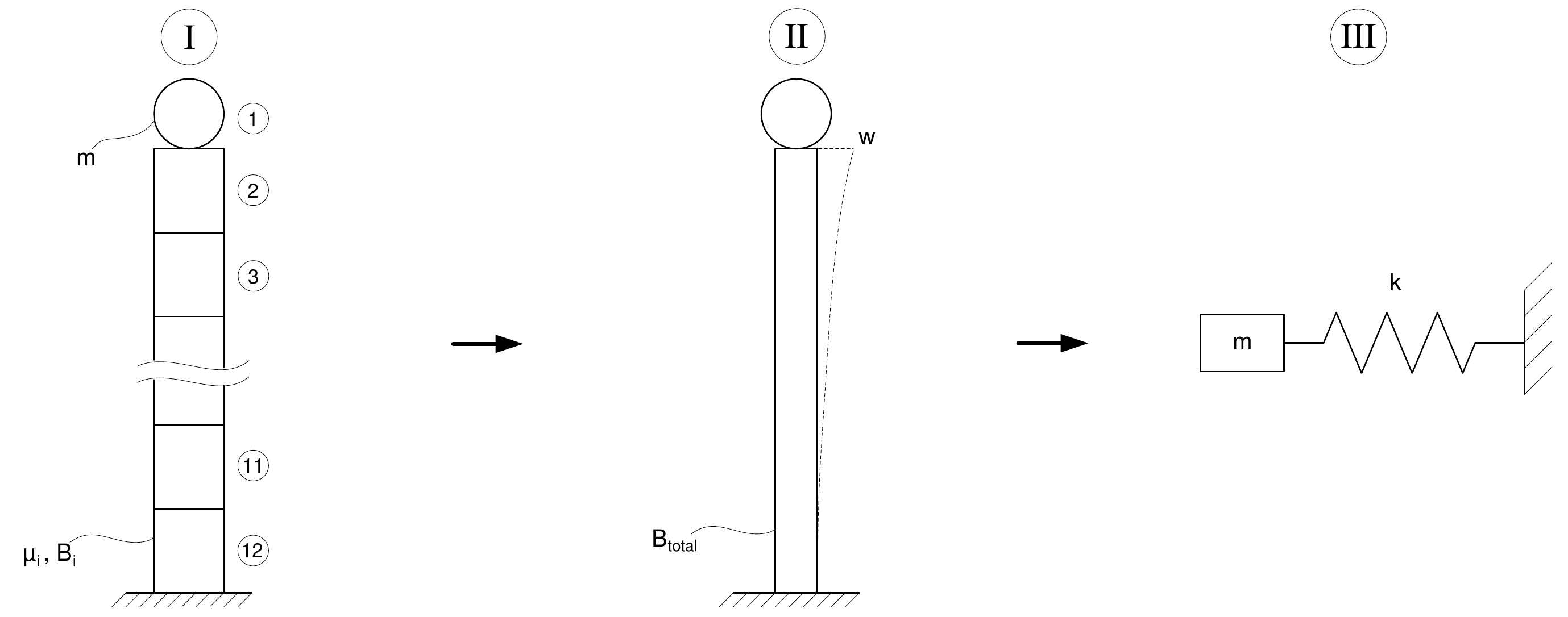}
\caption{Illustration of the direct stiffness method and the transition to a spring-mass system. I:
Tower model with several segments (11 are specified in \cite{Jonkman:2009}) and combined rotor-nacelle mass $m$; II: bending beam system with total
bending stiffness $B_{ges}$; III: spring-mass system with rotor-nacelle mass $m$ and translational stiffness $k$.}
\label{Fig_TurmBalkenFeder}
\end{figure}

The dynamics of fore-aft tower bending and flap-wise rotor blade bending are reduced to simple spring-mass-damper systems for the tower top and blade tip deflections. The respective tower stiffness coefficient $k_T$ is derived by means of a direct stiffness method common in structural mechanics calculations. The tower consisting of several segments is first transformed into an equivalent bending beam model. Afterwards, the bending stiffness of the effective beam model is transformed to an equivalent translational stiffness of the tower-nacelle dynamics (see Figure \ref{Fig_TurmBalkenFeder} and \ref{Sec_AppDirectStiffness}).
\newline While the tower stiffness parameter could be obtained and validated against the FAST simulation of the 5 MW reference turbine, there are still uncertainties about the determination of the blade parameter $k_B$, which is therefore adjusted according to FAST simulation results \cite{Georg:Fuzz2012}.

The effective mass $m_T$ for the tower-nacelle motion in equations \eqref{Eqn_SSMdl_4}, \eqref{Eqn_SSMdl_5}) is estimated as
\newline $m_T = m_\text{Rotor} + m_\text{Nacelle} + 0.25 \,m_\text{Tower}$, which has proven a reasonable assumption \cite{Gasch_eng:2012}. Similarly, the effective blade mass for the blade tip motion is estimated as $m_B = 0.25 \,m_\text{Blade}$.

\subsubsection{\label{Sec_AeroDamping}Aerodynamic Damping}

The aerodynamic rotor damping in fore-aft direction, which can be approximated as 
\newline $d_{11}\left(\lambda, \beta\right) = 0.5 \, \rho \, \pi \, R^2 \, v \, d_{11}^*\left(\lambda\right)$ \cite{Kaiser:2000}, is taken as an estimate for the damping parameter $d_T$ of the tower-top motion in equations \eqref{Eqn_SSMdl_4}, \eqref{Eqn_SSMdl_5}. The dimensionless parameter $d_{11}^*\left(\lambda,\beta\right)$ depends on the tip speed ratio and on the pitch angle and shows a similar behaviour for different turbine sizes \cite{Kaiser:2000}. Estimating $d_{11}\left(\lambda,\beta\right)$ accordingly for different stationary points of the 5 MW reference turbine for the whole operating range of the turbine yields values between
$3 \cdot 10^4 \,\frac{\text{Ns}}{\text{m}}$ and $10 \cdot 10^4 \, \frac{\text{Ns}}{\text{m}}$.
Here, the tower damping parameter is set to a constant value of $d_T = 7 \cdot 10^4 \,\frac{\text{Ns}}{\text{m}}$. The blade damping parameter is set to $d_B  = 2 \cdot 10^4 \,\frac{\text{Ns}}{\text{m}}$.

\subsubsection{\label{Sec_AeroMaps}Aero Maps}

The aero maps for the rotor thrust ($C_T$) and torque coefficients ($C_Q$) were extracted from FAST simulation runs of the 5 MW reference turbine.
Alternatively, they can be approximated using nonlinear functions \cite{Georg:Fuzz2012}.

\section{\label{Sec_TSObs}Observer in Takagi-Sugeno Model Structure}

In this section, a state-observer based on the nonlinear model \eqref{Eqn_StateSpaceMdl} is designed to reconstruct the unknown wind speed from the measurable system states. The standard Luenberg observer for linear systems is a state-space model including a feedback of the output error $\B{e}_\B{y} = \B{y} - \B{\yhat}$, where $\B{\yhat}$ is the reconstructed output signal:

\begin{equation}
 \B{\xhatP} \, = \, \B{A} \, \B{x} \, + \B{B \, u} \, + \, \B{L} \left(\B{y} - \B{\yhat}\right) \, , \qquad
 \B{\yhat}	 \, = \, \B{C \, \xhat} \, .
\label{Eqn_Luenberg_Obs}
\end{equation}
\newline As the wind turbine model is nonlinear, a linear observer like \eqref{Eqn_Luenberg_Obs} cannot be used in the whole operating range. Therefore, an observer in Takagi-Sugeno model structure is used.
\newline A state-space model in TS structure is of the form

\begin{equation}
 \B{\xP} \, = \, \sum\limits_{i=1}^{N_r} \, h_i(\B{z})\, \left(\B{A}_i \, \B{x} \, + \B{B}_i \, \B{u}\right) \, , \qquad
 \B{y}	 \, = \, \sum\limits_{i=1}^{N_r} \, h_i(\B{z})\, \B{C}_i \, \B{x} \, ,
\label{Eqn_ZustModel_TS}
\end{equation}
\newline where $\B{A}_i$, $\B{B}_i$ and $\B{C}_i$ are constant matrices and $h_i$ are nonlinear functions of the premise variables $\B{z}$, which can depend on the system states and inputs and on external variables. $N_r$ denotes the number of linear submodels. The membership functions $h_i$ fulfill the relation $\sum_{i=1}^{N_r} h_i = 1$. The linear submodels can be derived from the original nonlinear model using local Taylor linearisation or by applying the sector nonlinearity approach \cite{TanakaSano2:1994,Tanaka:2001}, whereby an exact representation of the nonlinear model is obtained. This approach is used in this paper for the derivation of the TS observer model.

\newpage

\subsection{\label{Sec_TSExample}Illustrating Example for a TS-Model}

A simple example shall be considered in order to illustrate the derivation of a TS model using sector nonlinearities.
\newline Consider the dynamic equation of a pendulum of length $l$ with a point mass $m$ driven by an external torque signal $M$:

\begin{equation}
\ddot{\varphi} \, = \, -\frac{g}{l}\,\sin\,\varphi \, + \, \frac{1}{m\,l^2} \, M \, ,
\label{Eqn_Pendulum}
\end{equation}
\newline where $\varphi$ denotes the angular displacement of the pendulum and $g$ the gravitational constant.

Introducing the state vector $\B{x} = \left(\varphi \quad \dot{\varphi}\right)^T$ and the input signal $u = M$, equation \eqref{Eqn_Pendulum} can be written in state-space form as

\begin{equation}
\B{\xP} \, = \, \begin{pmatrix}0	& 1 \\ -\frac{g}{l}\,\frac{\sin\,x_1}{x_1} & 0 \end{pmatrix}\,\B{x} \, + \, \begin{pmatrix} 0 \\ \frac{1}{m\,l^2} \end{pmatrix} \, u \,
			= \, \B{A}\left(\B{x}\right) \B{x}\, + \, \B{B}\,u \, .
\label{Eqn_Pendulum_StateSpace}
\end{equation}
\newline Obviously, this is a nonlinear model due to the function $f\left(x_1\right) = -\frac{g}{l}\,\frac{\sin\,x_1}{x_1}$. This function can be written as
\newline \(	f\left(x_1\right) \, = \, w_1\left(x_1\right) \, \fmax \, + \, w_2\left(x_1\right) \, \fmin \, , 
			\:\, \text{where} \:\: w_1\left(x_1\right) \,:= \,\Durch{f\left(x_1\right) - \fmin}{\fmax - \fmin} \, ,
			\: w_2\left(x_1\right) \, := \, \Durch{\fmax - f\left(x_1\right)}{\fmax - \fmin} \,. \)
\newline $\fmax$ and $\fmin$ denote the maximum and minimum values of the function $f$, i.e. the sector boundaries. However, any real constants $c_1$, $c_2$ could be used instead, as long as $c_1 \neq c_2$. Using the sector boundaries is advantageous, since the matrices of the linear submodels, which are used for TS controller and observer design, thereby contain the domain of the nonlinear system.

From the definition of $w_1$ and $w_2$ it is obvious that $w_1 + w_2 = 1$. Thus, the nonlinear matrix $\B{A}$ in \eqref{Eqn_Pendulum_StateSpace} can be written as

\begin{equation}
\B{A}\left(\B{x}\right) \, = \, \begin{pmatrix} 0 & w_1 + w_2 \\ w_1\,\fmax + w_2\,\fmin & 0 \end{pmatrix} \,
							= \, w_1\,\begin{pmatrix} 0 & 1 \\ \fmax  & 0 \end{pmatrix} \, + \, w_2\,\begin{pmatrix} 0 & 1 \\ \fmin  & 0 \end{pmatrix} \, 
							= \, w_1\,\B{A}_1 \, + \,w_2\, \B{A}_2 \, ,
\label{Eqn_TS_Pendulum}
\end{equation}
\newline and the whole model in \eqref{Eqn_Pendulum_StateSpace} as $\B{\xP} \, = \, \sum_{i=1}^2 \, w_i(x_1)\, \left(\B{A}_i \, \B{x} \, + \B{B} \, u\right)$.
\newline The nonlinearity has thus been shifted from the system matrix into the membership functions, which in this case are equivalent to the weighting functions $w_i$. In the same manner, systems with more than one nonlinearity can be transformed into a TS model structure by including all possible permutations of the $w_i$-functions into the membership functions $h_i$. The number $N_r$ of linear submodels generally is $N_r = 2^{N_l}$, where $N_l$ is the number of distinct nonlinear functions. However, if there occur several linear combinations of the same nonlinear function, $N_l$ is not increased.

\subsection{TS Observer}

The state-space model \eqref{Eqn_StateSpaceMdl} is used as a basis for the observer, where either the full model \eqref{Eqn_StateSpaceMdl} or submodels of \eqref{Eqn_StateSpaceMdl} can be used depending on the desired observer model order.

In this paper, only the rotational and torsional degrees of freedom are incorporated into the observer model but no tower and blade dynamics. This model configuration for the observer yields reasonable results while requiring relatively few measurement signals.
\newline In order to estimate the wind speed $v$ with a state observer, $v$ is included into the system state vector $\B{x}$ and a dynamic wind model is added to the system equations. The first-order delay model from \cite{Ekelund:1994} is used, modified by the mean value $\bar{v}$ of the wind speed, but without a white noise term:

\begin{equation}
\dot{v} \, = \, -\frac{1}{\tau_v} \,\left(v - \bar{v}\right)  \, ,
\label{Eq_v_mdl}
\end{equation}
\newline where the time constant is estimated as $\tau_v = 4\,\text{s}$. The mean wind speed $\bar{v}$ can be calculated over an appropriate time period (e.g. 10 min) from the anemometer wind measurement, which is sufficient for this purpose.

Since only the rotational and torsional degrees of freedom plus the estimated wind speed are considered for the observer model, the corresponding estimated state vector is
\newline $\B{\xhat} = \left(\hat{\theta}_s \quad \hat{\omega}_r \quad \hat{\omega}_g \quad \hat{v} \right)^T$. Since the first order pitch dynamics adds no information as to the reconstruction of the unknown states it is not considered in the observer model. This implies that the demanded pitch angle $\beta_d$ is not included in the input vector, because there is no linear dependence on $\beta_d$ but only a nonlinear dependence in $C_Q\left(\hat{\lambda}, \beta_d\right)$.
The mean wind speed $\bar{v}$ can be included in the input vector: $\B{u} = \left(T_g \quad \bar{v}\right)^T$.
The following states are assumed as measurable: $\theta_s$, $\omega_r$, $\omega_g$. Measuring the rotor and generator speed signals is routinely done in wind turbines. For a real application of the observer, the rotor speed signal would have to measured with high resolution and both speed signals might need to be filtered. It is only an assumption at this stage that the torsion angle is measurable. However, it should be possible to measure the rotation angles before and after the coupling between gearbox and generator (using for example absolute encoders), and thereby the torsion angle $\theta_s = \theta_r - \theta_g$, where the gearbox ratio can be taken into account simply as a factor.

From the system of nonlinear state-space equations for the wind turbine model \eqref{Eqn_StateSpaceMdl}, it is straightforward to obtain the nonlinear system matrix and the input matrix for the observer model:

\begin{equation}
\B{A}\left(\B{x}\right) \, = \, \begin{pmatrix}
			0       			&	1       			&	-1   				&	0									\\
            -\frac{k_S}{J_r}  	&	-\frac{d_S}{J_r}  	&   \frac{d_S}{J_r}		&	f\left(\B{\xhat},\,\beta_d\right)	\\
            \frac{k_S}{J_g}   	&   \frac{d_S}{J_g}   	&   -\frac{d_S}{J_g} 	&	0									\\
            0					&	0					&	0					&	-\frac{1}{\tau_v}
           \end{pmatrix} \, , \quad
\B{B} \, = \, \begin{pmatrix}
				0				&	0				\\
				0   			&	0				\\
				-\frac{1}{J_g}	&	0				\\
				0				&	\frac{1}{\tau_v}
			\end{pmatrix}	\, ,
\label{Eqn_NLMatrix_Obs1}
\end{equation}

$f\left(\B{\xhat},\,\beta_d\right) = \frac{1}{2\, J_r} \, \rho \, \pi \, R^3 \, \hat{v} \, C_Q\left(\hat{\lambda}, \beta_d\right) \, ,
	\qquad \fmin 	\, = \, 1.2414 \cdot 10^{-5} \, \frac{1}{\text{m s}} \, , \qquad \fmax 	\, = \, 0.0559 \, \frac{1}{\text{m s}}$

The values for $\fmin$ and $\fmax$ were obtained by estimating the minimum and maximum values of the wind speed $v$ and the torque coefficient $C_Q$: $C_{Q,\text{max}} = 0.0751$,  $C_{Q,\text{min}} = 0.001$, $v_\text{max} = 60 \,\frac{\text{m}}{\text{s}}$, $v_\text{min} = 1 \,\frac{\text{m}}{\text{s}}$. Though $v_\text{min}$ and $C_{Q,\text{min}}$ are zero in theory, they are set to small positive values to avoid generating zero entries in one of the TS submatrices.
The output vector and the output matrix are given by $\B{y} = \left(\theta_s \quad \omega_r \quad \omega_g\right)^T$ and $\B{C} \, = \, \left(\B{I}_{3 \x 3} \: \B{0}_{3 \x 1}\right)$.

Employing the same procedure as in section \ref{Sec_TSExample}, the observer model can be obtained in TS structure:

\begin{equation}
 \B{\xhatP} \, = \, \sum\limits_{i=1}^{N_r\,=\,2} \, h_i(\B{\zhat})\, \left(\B{A}_i \, \B{\xhat} \, + \B{B \, u} \, + \, \B{L}_i \left(\B{y} - \B{\yhat}\right) \right) \, , \qquad
 \B{\yhat}	 \, = \, \B{C \, \xhat} \, ,
\label{Eqn_TS_Obs}
\end{equation}
\newline where the premise variable $\zhatB$ now depends on the reconstructed states: $\zhatB = \left(\hat{\omega}_r \:\: \hat{v} \:\: \beta_d\right)^T$.

\subsection{\label{Sec_Gains_and_Stability}Observer Gains and Stability}

A common means to derive gain matrices for observers in TS structure is by applying the direct method of Lyapunov in form of linear matrix inequalities (LMI) \cite{Lendek:2010}.

In general, the global asymptotic stability of a nonlinear system $\B{\xP} = \B{f}\left(\B{x}\right)$ is guaranteed if there exists a Lyapunov function $V\left(\B{x}\right)$ satisfying the conditions $V\left(\B{x}\right) > 0$ and $\dot{V}\left(\B{x}\right) < 0$ for all trajectories. In particular, the system is stable if it is quadratically stable, i.e., if a quadratic Lyapunov function $V = \B{x}^T\B{P}\B{x}$, with a symmetric, positive definite matrix $\B{P}$, exists.
\newline In that case, for a TS system without an external input ($\B{\xP} = \sum_{i=1}^{N_r} h_i\left(\B{z}\right)\, \B{A}_i$), the condition $\dot{V}\left(\B{x}\right) < 0$ is equivalent to
$ \dot{V} =  \B{\xP}^T\,\B{P}\,\B{x} \, + \, \B{x}^T\,\B{P}\,\B{\xP} = 
					\B{x}^T \, \left(\sum_{i=1}^{N_r} h_i\left(\B{z}\right)\,\left( \B{A}^T_i \,\B{P} \, + \, \B{P}\,\B{A}_i \right) \right) \, \B{x} \, < \, 0 $. Since this condition must hold for all $\B{x}$, the TS system without input is stable if there exists a common symmetric, positive definite matrix $\B{P}$, such that

\begin{equation}
 \B{A}^T_i \,\B{P} \, + \, \B{P}\,\B{A}_i \, < \, 0 \qquad \left(i \in \left\lbrace 1,\dots , N_r\right\rbrace \right) \, . \quad \cite{TanakaSugeno:1992, WangTanakaGriffin:1996}
 \label{Eqn_LyapTS}
\end{equation}
\newline For the TS observer \eqref{Eqn_TS_Obs}, where the membership functions depend on unmeasurable states ($h_i = h_i\left(\widehat{\B{z}}\right)$), a modified form of the stability condition \eqref{Eqn_LyapTS} with an additional LMI can be used to guarantee the stability of the error dynamics of the observer system \cite{BergstenPalmDriankov:2001}:

\begin{equation}
\B{P}\, \left(\B{A}_i - \B{L}_i \B{C}\right) + \left(\B{A}_i - \B{L}_i \B{C}\right)^T \, \B{P} \leq - \B{Q} \, , \qquad
\begin{pmatrix}
\B{Q}\ - \mu^2 I	& \B{P} \\
\B{P}				& \B{I}
\end{pmatrix} \, > \, 0 \, , 	\label{Eqn_LyapTSObs_est}
\end{equation}
\newline where $\B{Q}$ is a symmetric, positive definite matrix and $\mu >0$ is a known constant satisfying $\Delta\left(\B{z},\zhatB\right) \leq \mu \|\B{e}\|$,
with $\B{e} = \|\B{x}-\B{\xhat}\|$ and 
$\Delta\left(\B{z},\zhatB\right) \, = \, \left\|\sum_{i=1}^{N_r} \left(h_i(\B{z}) - h_i(\zhatB)\right) \left(\B{A}_i \B{x} + \B{B} \B{u}\right)\right\|$. The first inequality of \eqref{Eqn_LyapTSObs_est} is not an LMI but can be recast in LMI form by introducing $\B{N}_i := \B{P} \B{L}_i$ \cite{Tanaka:2001}. As condition \eqref{Eqn_LyapTSObs_est} concerns quadratic stability, it is only a sufficient stability condition, i.e., if it is not fulfilled, no formal statement can be made about the stability or instability of the considered system \cite{Lendek:2010}.

\subsubsection*{Optimal LMI Observer Design}

Condition \eqref{Eqn_LyapTSObs_est} was first used to calculate the observer gains. However, this observer hardly had any modifying effect on the wind speed compared to the mere wind model \eqref{Eq_v_mdl}.
A possible remedy is to modify the gain matrices with a weighting matrix, such that the gains influencing the wind speed $\hat{v}$ are increased. A more systematic way is to make use of optimal fuzzy control concepts, where weighting matrices for the system states/outputs and inputs and a quadratic cost function can be included in the LMIs \cite{TanakaTaniguchiWang1:1998}.
\newline For the observer design in this paper, theorem 5 from \cite{TanakaTaniguchiWang1:1998}, which is applicable for controller design, was modified to be used for the dual TS-systems $\left(\B{A}_i^T, \B{C}^T\right)$. The observer gain matrices $\B{L}_i$ are then obtained from the resulting gain matrices $\B{K}_i$ as $\B{L}_i = \B{K}_i^T$. The obtained gain matrices are given in \ref{Sec_ObserverGains}, \eqref{Eqn_L_Matrices}.

The formal stability of the error dynamics could not be verified with condition \eqref{Eqn_LyapTSObs_est}, which is a conservative condition due to the assumption of un unstructured uncertainty \cite{Lendek:2010}. However, the observer shows a stable behaviour in the FAST simulation, even for large initial observer errors (see section \ref{Sec_Simulation}).

\section{\label{Sec_Simulation}Simulation Results}

The observer was integrated in the FAST / Simulink{\regsym} model and simulated using an IEC wind gust as well as turbulent wind input. To control the rotor speed, a state-space controller in TS structure, based on Taylor-linearised models was used.

\subsubsection{Simulation with IEC Wind Gust}

\begin{figure}[htbp!]
\centerline{\subfloat{\includegraphics[width=0.8\linewidth]{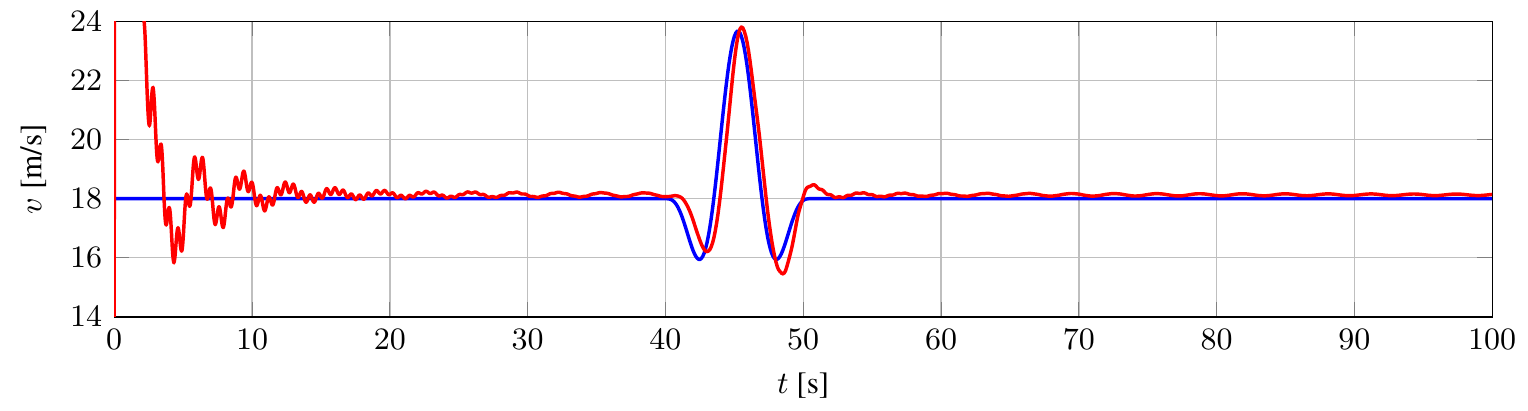}}}
\vfil
\subfloat{\includegraphics[width=0.32\linewidth]{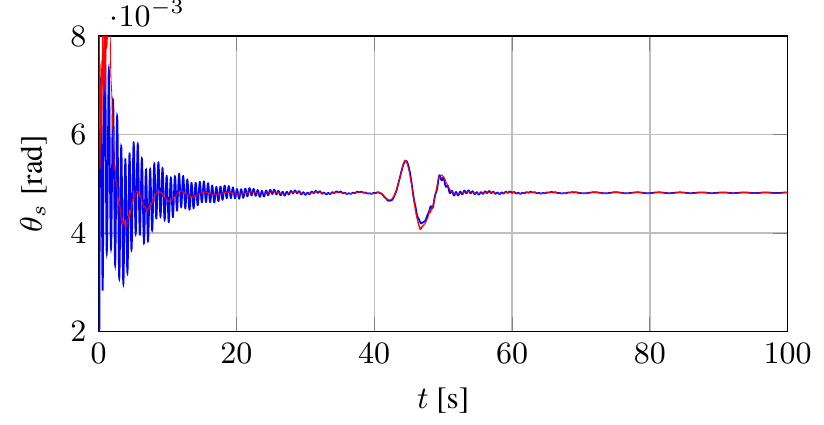}}
\hfil
\subfloat{\includegraphics[width=0.32\linewidth]{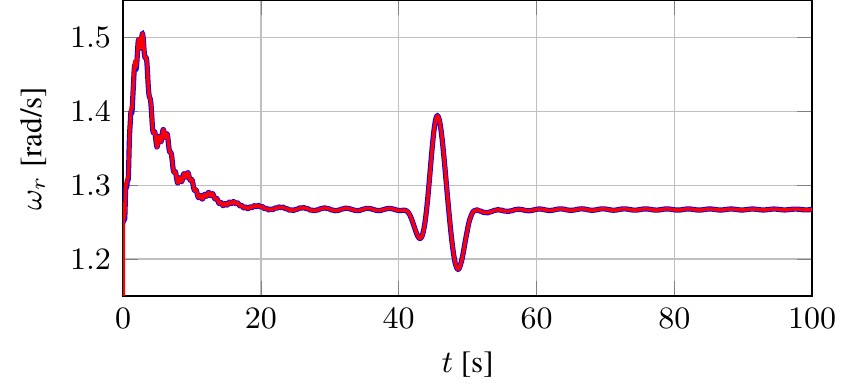}}
\hfil
\subfloat{\includegraphics[width=0.32\linewidth]{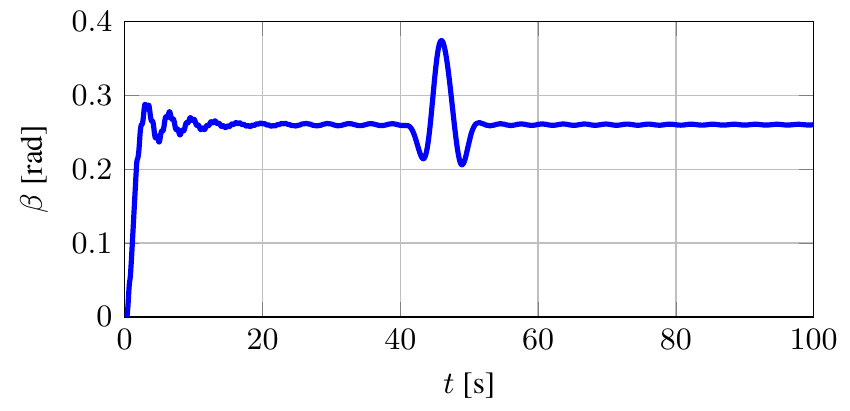}}
\caption{Simulation with IEC wind gust. \textcolor{blue}{Blue}: True states from FAST simulation; \textcolor{red}{red}: Estimated states; Initial values:
			$\theta_{s,0} = 0\,\text{rad}$, $\hat{\theta}_{s,0} = 0.1\,\text{rad}$, $\omega_{r,0} = \omega_{g,0} = 1.267 \,\frac{\text{rad}}{\text{s}}$,
			$\hat{\omega}_{r,0} = \hat{\omega}_{g,0} = 0 \,\frac{\text{rad}}{\text{s}}$, $v_0 = 18 \,\frac{\text{m}}{\text{s}}$,
			$\hat{v}_0 = 1 \,\frac{\text{m}}{\text{s}}$. {The torsion angle is not directly available from the FAST outputs and was obtained by
			integrating the speed error signal from the FAST outputs of rotor and generator speed (corrected by the gear ratio). The pitch angle is only shown for reference.}}
\label{Figs_obs_FAST}
\end{figure}

As a first simulation test, an IEC wind gust with mean wind speed $18\,\frac{\text{m}}{\text{s}}$ was used. Results are shown in figure \ref{Figs_obs_FAST}. 
After the transient, the observer shows a stable behaviour and the wind gust is reconstructed with a phase offset of $\approx 0.4 - 0.5$ s. Almost perfect reconstruction is achieved for the rotor speed $\omega_r$.

\subsubsection{Simulation with Turbulent Wind}

\begin{figure}[htbp!]
\centerline{\subfloat{\includegraphics[width=0.8\linewidth]{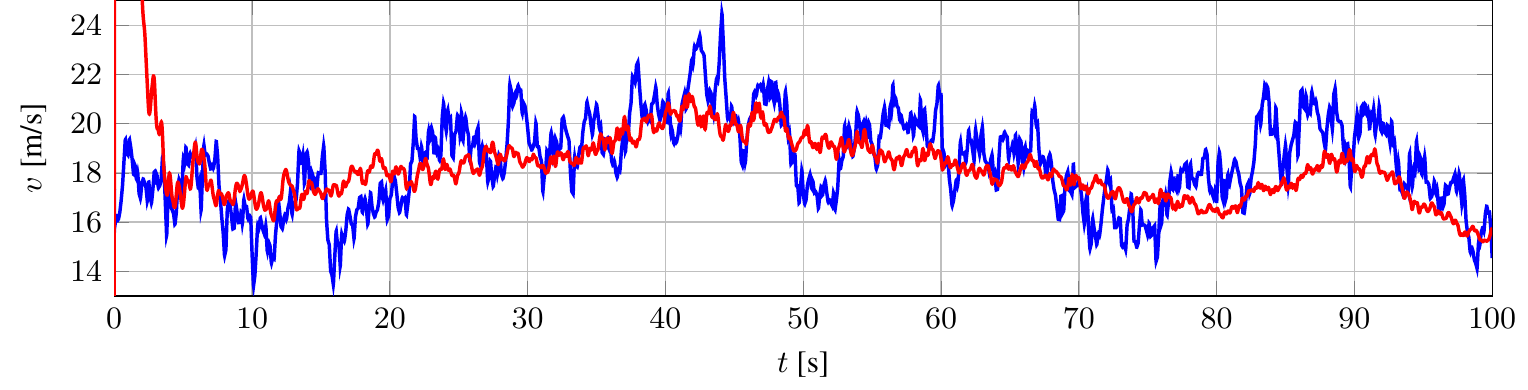}}}
\vfil
\subfloat{\includegraphics[width=0.32\linewidth]{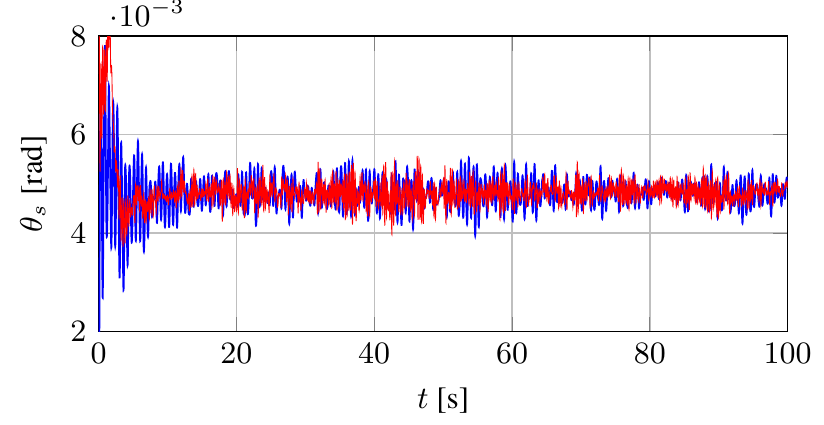}}
\hfil
\subfloat{\includegraphics[width=0.32\linewidth]{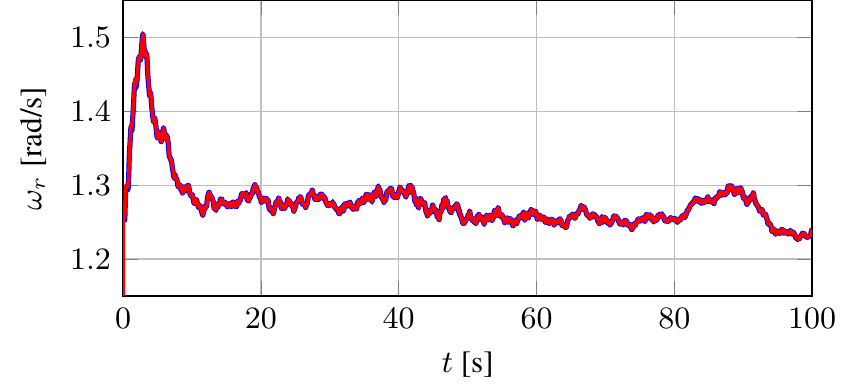}}
\hfil
\subfloat{\includegraphics[width=0.32\linewidth]{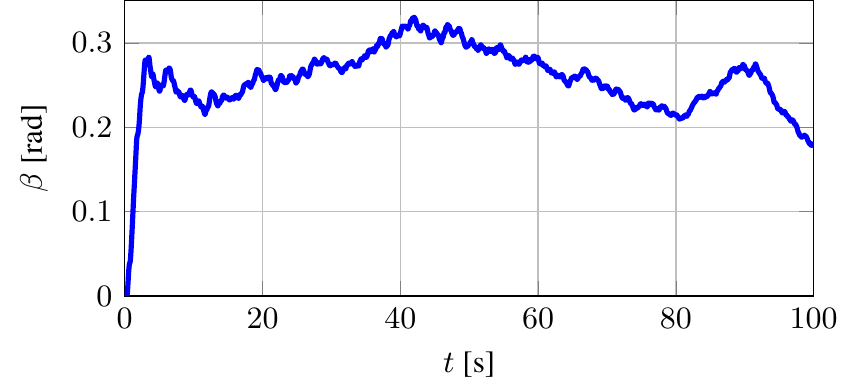}}
\caption{Turbulent wind simulation results. \textcolor{blue}{Blue}: True states from FAST simulation; \textcolor{red}{red}: Estimated states; Initial values:
			$\theta_{s,0} = 0\,\text{rad}$, $\hat{\theta}_{s,0} = 0.1\,\text{rad}$, $\omega_{r,0} = \omega_{g,0} = 1.267 \,\frac{\text{rad}}{\text{s}}$,
			$\hat{\omega}_{r,0} = \hat{\omega}_{g,0} = 0 \,\frac{\text{rad}}{\text{s}}$,
			$\hat{v}_0 = 1 \,\frac{\text{m}}{\text{s}}$. The \textcolor{blue}{blue} wind speed signal from FAST is the nominal downwind component of the hub-height wind speed, not the rotor effective wind speed.}
\label{Figs_obs_FAST_turb}
\end{figure}

A second simulation run in FAST was done using a 3D turbulent wind field with mean wind speed $18\,\frac{\text{m}}{\text{s}}$. Results are shown in figure
\ref{Figs_obs_FAST_turb}. When interpreting figure \ref{Figs_obs_FAST_turb}, it is important to remember that the observer estimates the rotor effective wind speed, i.e. a virtual single point wind speed that causes the same variations in wind torque as the corresponding 3D turbulent wind field \cite{DOWECreport}. Although the calculations in FAST are based on the 3D wind field, the wind speed output from FAST (\textcolor{blue}{blue} curve in figure \ref{Figs_obs_FAST_turb}) shows the nominal downwind component of the hub-height wind speed, so the two wind speed curves in figure \ref{Figs_obs_FAST_turb} are not directly comparable. The FAST wind speed output is shown to give an idea of the observer performance.

In case the observer shall be tested in a real turbine, the estimated effective wind speed would not be directly comparable either to single point measurements on the nacelle or on meteorological towers. It would be interesting to compare the estimated wind speed to LIDAR measurements of the wind field measured in front of the rotor.

\section{\label{Sec_Conclusion}Conclusion and Outlook}

In this paper, a nonlinear observer in Takagi-Sugeno structure was designed to estimate the effective wind speed from the measurable states of a dynamic wind turbine model. Although formal stability of the observer in terms of LMI conditions could not be obtained, the observer shows a stable behaviour when used with the aero-elastic simulation code FAST.

The TS observer for wind speed estimation is intended as one module of a fault-tolerant control scheme for wind turbines in future work.

\subsubsection*{Acknowledgement}

This work was conducted within a research project funded by the German Federal Ministry of Education and Research under grant no. 17N1411.

\appendix

\section{\label{Sec_AppDirectStiffness}Derivation of Effective Tower Stiffness}

The direct stiffness method allows to calculate eigenfrequencies and eigenmodes of structures consisting of several segments of defined length, mass and bending stiffness. For each segment, the characteristic forces and displacements can be calculated from the previous segment by means of a transfer matrix depending on the frequency of the structure \cite{GaschKnothe2}. Applying the total transfer matrix as the product of the individual transfer matrices, as well as the boundary conditions for the rigid and the free ends of the beam, yields a homogeneous system of equations for the displacements at the top of the total structure, which is fulfilled for the eigenfrequencies of the structure.
In order to calculate the respective equivalent bending stiffness, it is sufficient to find the first eigenfrequency $\omega_1$. For the tower, it was calculated as $\omega_1 \approx 2.14 \, \rads$ and has been validated with the NREL-Software BModes \cite{BModes} ($\omega_{1,\text{BModes}} \approx 2.08 \, \rads $). The connection to the equivalent bending stiffness $B_{total}$ is
\begin{equation}
\omega_1 \, = \, \kappa_1^2 \; \sqrt{\frac{B_{total}}{\mu_{total}}} \quad \Rightarrow \quad B_{total} \,
			= \, \frac{\omega_1^2 \;  \mu_{total}}{\kappa_1^4} \, \approx \, 4.44 \cdot 10^{11} \, \text{Nm}^2 \, ,
\label{gl:B_total}
\end{equation}
where $\kappa_1 = 1.423 \cdot 10^{-2}$ is a factor that can be found in standard mechanics textbooks and $\mu_{total}$ is the total mass per length.
Finally, the equivalent bending stiffness can be transferred into a translational spring stiffness with simple equations for the deflection $w$ of the beam (with total length $l$) and spring, where the applied force $F$ corresponds to the rotor thrust force $F_T$:
\begin{equation}
w=\frac{F\;l^3}{3 B_{total}} \quad (beam), \; \; F=k \, w \quad (spring) \qquad \Rightarrow \quad k=\frac{3B_{total}}{l^3} \approx 1.98 \cdot 10^6 \, \frac{\text{N}}{\text{m}}
\end{equation}

\section{\label{Sec_AppModelParameters}Model Parameters}

$N = 3$, $R = 63$ m, $\rho = 1.225 \, \frac{\text{kg}}{\text{m}^3}$, $J_r = 38759227 \, \text{kg}\,\text{m}^2$,
$J_g = 5025347 \, \text{kg}\,\text{m}^2$
\newline $k_s = 867637000 \, \text{Nm}$, $d_s = 6215000 \, \text{Nm\,s}$, $k_B = 40000 \, \frac{\text{N}}{\text{m}}$, $\alpha = 0.02 \, \text{m}^{-1}$, 
$k_T = 1.98 \cdot 10^6 \, \frac{\text{N}}{\text{m}}$
\newline $m_\text{Blade} = 17740 \, \text{kg}$, $m_\text{Tower} = 347640 \, \text{kg}$, $m_\text{Rotor} = 110000 \, \text{kg}$, $m_\text{Nacelle} = 240000 \, \text{kg}$,
\newline $m_T = 436865 \,\text{kg}$, $m_B = 4435 \,\text{kg}$, $d_T = 7 \cdot 10^4 \frac{\text{Ns}}{\text{m}}$, $d_B = 2 \cdot 10^4 \frac{\text{Ns}}{\text{m}}$,
$\tau = 0.1 \,\text{s}$, $\tau_v = 4 \,\text{s}$, $r_B = 21.975 \,\text{m}$

\section{\label{Sec_ObserverGains}Observer Gain Matrices}

The following weighting matrices ($\B{W}$ for the system states and $\B{R}$ for the system outputs) were used for the optimal LMI observer design:
\newline $\B{W} = \text{diag}\left(\frac{W_1}{\theta^2_{s,\text{max}}} \, , \quad \frac{W_2}{\omega^2_{r,\text{max}}} \, , \quad
							\frac{W_3}{\omega^2_{g,\text{max}}} \, , \quad \frac{W_4}{v^2_\text{max}} \right)$,
$\B{R} = \text{diag}\left(\frac{R_1}{\theta^2_{s,\text{max}}} \, , \quad \frac{R_2}{\omega^2_{r,\text{max}}} \, , \quad \frac{R_3}{\omega^2_{g,\text{max}}}\right)$ , 
\newline with $W_1    = 0.25$, $W_2    = 15.708$, $W_3    = 1.5708$, $W_4    = 60 \cdot 10^7$, $R_1    = 0.05$, $R_2    = 0.1571$, $R_3    = 1.5708$ and the estimated maximum values
\newline $\theta_{s,\text{max}} = 0.01 \, \text{rad}$, 
$\omega_{r,\text{max}} = \omega_{g,\text{max}} = 15\,\cdot \frac{\pi}{30} \, \frac{\text{rad}}{\text{s}}$,
$v_\text{max} = 60\, \frac{\text{m}}{\text{s}}$ to normalise the chosen weights.
\newline For the optimal LMI design procedure, the initial observer error is needed, which was set to $\B{e}_0=\left(0 \: 0 \: 0 \: 0\right)^T$. This is of course an idealisation. However, as can be seen from the simulation results, the observer is stable also for $\|\B{e}_0\| > 0$.
\newline The following observer gain matrices were obtained:
\begin{equation}
L_1 \, = \, \begin{pmatrix}
				0.147  		&	-176.5		&	143.6	\\
				-0.022		&	133			&	-28.6	\\
				0.183		&	-286.1		&	303.2	\\
				0.08		&	6698.1		&	741.2
			\end{pmatrix} \, , \qquad
L_2 \, = \, \begin{pmatrix}
				0.147  		&	-176.5		&	143.6	\\
				-0.022		&	133			&	-28.6	\\
				0.183		&	-286.1		&	303.2	\\
				0.08		&	6698.1		&	741.2
			\end{pmatrix}
\label{Eqn_L_Matrices}
\end{equation}
\newline $\B{L}_1$ and $\B{L}_2$, displayed here with rounded values, are not equal but differ by less than 0.1 \%.

\bibliographystyle{plain}
\bibliography{Literature}

\begin{thebibliography}{10}

\bibitem{BergstenPalmDriankov:2001}
Pontus Bergsten, R.~Palm, and D.~Driankov.
\newblock Fuzzy {O}bservers.
\newblock In {\em IEEE International Conference on Fuzzy Systems}, pages
  700--703, Melbourne, Australia, 2001.

\bibitem{Bianchi:2007}
Fernando~D. Bianchi, Hern\'{a}n {De Battista}, and Ricardo~J. Mantz.
\newblock {\em Wind Turbine Control Systems - Principles, Modelling and Gain
  Scheduling Design}.
\newblock Springer-Verlag, London Limited, 2007.

\bibitem{Bindner:1999}
Henrik Bindner.
\newblock Active {C}ontrol: {W}ind {T}urbine {M}odel.
\newblock Technical report, Ris{\o}-R-920(EN), Ris{\o} National Laboratory,
  Roskilde, Denmark, 1999.

\bibitem{BModes}
Gunjit Bir.
\newblock {\em {NWTC} {D}esign {C}odes ({BM}odes by {G}unjit {B}ir).
  http://wind.nrel.gov/designcodes/preprocessors/bmodes/}.
\newblock NREL, 2012.

\bibitem{Ekelund:1994}
Thommy Ekelund.
\newblock Speed {C}ontrol of {W}ind {T}urbines in the {S}tall {R}egion.
\newblock In {\em IEEE Conference on Control Applications}, pages 227 -- 232,
  Glasgow, UK, 1994.

\bibitem{Gasch_eng:2012}
Robert Gasch and Jochen~Twele (eds.).
\newblock {\em Wind Power Plants}.
\newblock Springer-Verlag, Berlin, Heidelberg, 2nd edition, 2012.

\bibitem{GaschKnothe2}
Robert Gasch and Klaus Knothe.
\newblock {\em Strukturdynamik, Band 2: Kontinua und ihre Diskretisierung}.
\newblock Springer-Verlag Berlin, Heidelberg, 1989.

\bibitem{Georg:Fuzz2012}
S\"oren Georg, Horst Schulte, and Harald Aschemann.
\newblock Control-{O}riented {M}odelling of {W}ind {T}urbines {U}sing a
  {T}akagi-{S}ugeno {M}odel {S}tructure.
\newblock In {\em IEEE International Conference on Fuzzy Systems}, pages
  1737--1744, Brisbane, Australia, 2012.

\bibitem{GerlandSchulte:2010}
Patrick Gerland, Dominic Gro{\ss}, Horst Schulte, and Andreas Kroll.
\newblock Design of {S}liding {M}ode {O}bservers for {TS} {F}uzzy {S}ystems
  with {A}pplication to {D}isturbance and {A}ctuator {F}ault {E}stimation.
\newblock In {\em IEEE Conference on Decision and Control}, pages 4373--4378,
  Atlanta, USA, 2010.

\bibitem{Jonkman:2009}
J.~Jonkman, S.~Butterfield, W.~Musial, and G.~Scott.
\newblock Definition of a 5-{MW} {R}eference {W}ind {T}urbine for {O}ffshore
  {S}ystem {D}evelopment.
\newblock Technical report, NREL/TP-500-38060, National Renewable Energy
  Laboratory, Golden, Colorado, 2009.

\bibitem{FASTUserGuide}
Jason~M. Jonkman and Marshall~L. {Buhl Jr.}
\newblock {FAST} {U}ser's {G}uide.
\newblock Technical report, NREL/EL-500-38230, National Renewable Energy
  Laboratory, Golden, Colorado, 2005.

\bibitem{Kaiser:2000}
Klaus Kaiser.
\newblock {\em Luftkraftverursachte {S}teifigkeits- und {D}\"ampfungsmatrizen
  von {W}indturbinen und ihr {E}influ{\ss} auf das {S}tabilit\"atsverhalten}.
\newblock VDI-Fortschritt-Berichte, Nr. 294, VDI-Verlag D\"usseldorf, 2000.

\bibitem{Lendek:2010}
Zs\'{o}fia Lendek, Thierry~Marie Guerra, Robert Babu\v{s}ka, and Bart {De
  Schutter}.
\newblock {\em Stability Analysis and Nonlinear Observer Design Using
  Takagi-Sugeno Fuzzy Models}.
\newblock Springer-Verlag Berlin Heidelberg, 2010.

\bibitem{MaPoulsen_EstWind:1995}
Xin Ma, Niels~K. Poulsen, and H.~Bindner.
\newblock Estimation of {W}ind {S}peed in {C}onnection to a {W}ind {T}urbine.
\newblock Technical report, The Technical University of Denmark, 1995.

\bibitem{Ostergaard_v_est:2007}
K.~Z. {\O}stergaard, P.~Brath, and J.~Stoustrup.
\newblock Estimation of effective wind speed.
\newblock {\em Proc. The Science of Making Torque from Wind, J. Phys.: Conf.
  Ser.}, 75(012082), 2007.

\bibitem{SchulteZajac:2012}
Horst Schulte, Michal Zajac, and S\"oren Georg.
\newblock {T}akagi-{S}ugeno {S}liding {M}ode {O}bserver {D}esign for {L}oad
  {E}stimation and {S}ensor {F}ault {D}etection in {W}ind {T}urbines.
\newblock In {\em IEEE International Conference on Fuzzy Systems}, pages
  292--299, Brisbane, Australia, 2012.

\bibitem{SugenoKang:1988}
M.~Sugeno and G.~T. Kang.
\newblock Structure {I}dentification of {F}uzzy {M}odel.
\newblock {\em Fuzzy Sets and Systems}, 28:15--33, 1988.

\bibitem{TakagiSugeno}
T.~Takagi and M.~Sugeno.
\newblock Fuzzy {I}dentification of {S}ystems and {I}ts {A}pplication to
  {M}odeling and {C}ontrol.
\newblock {\em IEEE Transactions on Systems, Man, and Cybernetics},
  15(1):116--132, 1985.

\bibitem{TanakaSano2:1994}
K.~Tanaka and M.~Sano.
\newblock A {R}obust {S}tabilization {P}roblem of {F}uzzy {C}ontrol {S}ystems
  and {I}ts {A}pplication to {B}acking up {C}ontrol of a {T}ruck-{T}railer.
\newblock {\em IEEE Transactions on Fuzzy Systems}, 2(2):119--134, 1994.

\bibitem{TanakaTaniguchiWang1:1998}
K.~Tanaka, T.~Taniguchi, and H.~O. Wang.
\newblock Fuzzy {C}ontrol {B}ased on {Q}uadratic {P}erformance {F}unction - {A}
  {L}inear {M}atrix {I}nequality {A}pproach.
\newblock In {\em IEEE Conference on Decision and Control}, pages 2914--2919,
  Tampa, USA, 1998.

\bibitem{TanakaSugeno:1992}
Kazuo Tanaka and Micheo Sugeno.
\newblock Stability analysis and design of fuzzy control systems.
\newblock {\em Fuzzy Sets and Systems}, 45(2):135--156, 1992.

\bibitem{Tanaka:2001}
Kazuo Tanaka and Hua~O. Wang.
\newblock {\em Fuzzy Control Systems Design and Analysis: A Linear Matrix
  Inequality Approach}.
\newblock John Wiley \& Sons, Inc., 2001.

\bibitem{DOWECreport}
E.~L. {van der Hooft}, P.~Schaak, and T.~G. {van Engelen}.
\newblock Wind {T}urbine {C}ontrol {A}lgorithms.
\newblock Technical report, DOWEC-F1W1-EH-03-094/0, ECN-C--03-111, 2003.

\bibitem{WangTanakaGriffin:1996}
H.~O. Wang, K.~Tanaka, and M.~F. Griffin.
\newblock An {A}pproach to {F}uzzy {C}ontrol of {N}onlinear {S}ystems:
  {S}tability and {D}esign {I}ssues.
\newblock {\em IEEE Transactions on Fuzzy Systems}, 4(1):14--23, 1996.

\end{thebibliography}

\end{document}